\documentclass[iop]{emulateapj}
\usepackage[utf8]{inputenc} 
\usepackage[letterpaper,pass]{geometry}
\usepackage{apjfonts}

\usepackage[usenames,dvipsnames]{color}
\usepackage{amsmath}
\usepackage[unicode,breaklinks]{hyperref}
\usepackage[all]{hypcap}
\hypersetup{backref,
            bookmarks,
            dvips,
            pdftitle=C:,
            pdfsubject=Astrophysics \& Cosmology,
            pdfkeywords={},
            pdfdisplaydoctitle,
            colorlinks=true,
            citecolor=blue,
            pdfauthor=Yohai Meiron,
            unicode}

\newcommand{\lmax}{l_\mathrm{max}}
\newcommand{\nmax}{n_\mathrm{max}}

\usepackage{etoolbox}

\makeatletter

\patchcmd{\NAT@citex}
  {\@citea\NAT@hyper@{%
     \NAT@nmfmt{\NAT@nm}%
     \hyper@natlinkbreak{\NAT@aysep\NAT@spacechar}{\@citeb\@extra@b@citeb}%
     \NAT@date}}
  {\@citea\NAT@nmfmt{\NAT@nm}%
   \NAT@aysep\NAT@spacechar\NAT@hyper@{\NAT@date}}{}{}

\patchcmd{\NAT@citex}
  {\@citea\NAT@hyper@{%
     \NAT@nmfmt{\NAT@nm}%
     \hyper@natlinkbreak{\NAT@spacechar\NAT@@open\if*#1*\else#1\NAT@spacechar\fi}%
       {\@citeb\@extra@b@citeb}%
     \NAT@date}}
  {\@citea\NAT@nmfmt{\NAT@nm}%
   \NAT@spacechar\NAT@@open\if*#1*\else#1\NAT@spacechar\fi\NAT@hyper@{\NAT@date}}
  {}{}

\makeatother

\slugcomment{Submitted to the Astrophysical Journal}

\shorttitle{Expansion Techniques}
\shortauthors{Meiron et al.}

\begin{document}

\title{Expansion Techniques for Collisionless Stellar Dynamical Simulations}

\author{Yohai Meiron\altaffilmark{1,6}, Baile Li\altaffilmark{2}, Kelly Holley-Bockelmann\altaffilmark{2,3}, and Rainer Spurzem\altaffilmark{4,1,5}}
\affil{
    \altaffilmark{1}{Kavli Institute for Astronomy and Astrophysics at Peking University, Beijing 100871, China}\\
    \altaffilmark{2}{Department of Physics and Astronomy, Vanderbilt University, Nashville, TN 37235, USA}\\
    \altaffilmark{3}{Department of Physics, Fisk University, Nashville, TN 37208, USA}\\
    \altaffilmark{4}{National Astronomical Observatories of China, Chinese Academy of Sciences, Beijing 100012, China}\\
    \altaffilmark{5}{Astronomisches Rechen-Institut, Zentrum für Astronomie, Universität Heidelberg, Heidelberg D-69120, Germany}\\
}
\altaffiltext{6}{email: \href{mailto:ymeiron@pku.edu.cn}{ymeiron@pku.edu.cn}}

\begin{abstract}
We present GPU implementations of two fast force calculation methods, based on series expansions of the Poisson equation. One is the Self-Consistent Field (SCF) method, which is a Fourier-like expansion of the density field in some basis set; the other is the Multipole Expansion (MEX) method, which is a Taylor-like expansion of the Green's function. MEX, which has been advocated in the past, has not gained as much popularity as SCF. Both are particle-field method and optimized for collisionless galactic dynamics, but while SCF is a ``pure'' expansion, MEX is an expansion in just the angular part; it is thus capable of capturing radial structure easily, where SCF needs a large number of radial terms. We show that despite the expansion bias, these methods are more accurate than direct techniques for the same number of particles. The performance of our GPU code, which we call \textit{ETICS}, is profiled and compared to a CPU implementation. On the tested GPU hardware, a full force calculation for one million particles took $\sim 0.1$ seconds (depending on expansion cutoff), making simulations with as many as $10^8$ particles fast on a comparatively small number of nodes.
\end{abstract}


\section{Introduction}

A galaxy is a self-gravitating system where stellar dynamics is governed by Newton's law. It could be naively described as a set of $3N_\star$ coupled, second-order, non-linear ordinary differential equations, where $N_\star$ is the number of stars, which ranges between $10^5$ and $10^{12}$ \citep{BinneyTremaine}. Solving such an equation set numerically is practically only possible at the very low end of the $N_\star$-range, and even so very challenging with current computer hardware. Thus, various techniques are used to simplify the mathematical description of the system; these are often designed to fit a particular problem in stellar dynamics and yield unphysical results when applied to another problem.

Direct $N$-body simulation is one of the main techniques used to study gravitational systems in general and galaxies in particular. In this technique, the distribution function is sampled at $N \ll N_\star$ points in a Monte-Carlo fashion. This $N$ depends on the computational capabilities, and an astrophysical system with $10^{11}$ stars might be represented numerically by a sample of just $10^5$ ``supermassive'' particles. This seems to be allowed because of the equivalence principle and the fact that gravitation is scale free, unlike, for example, in molecular dynamics. However in gravity too this simplification can cause problems, as some dynamical effects depend on number density rather than just mass density.

The most well known $N$-dependent effect in stellar dynamics is two-body relaxation. The relaxation time, the characteristic time for a particle's velocity to change by order of itself due to encounters with other particles, scales with the crossing time roughly as $N/\ln N$. Thus, the ratio between the relaxation times in a real and a simulated system is of similar order of magnitude to the undersampling factor. This could be taken into account when interpreting the result of an undersampled simulation, but a poorly sampled distribution function might have other, unexpected, consequences.

Galaxies are often described as collisionless stellar systems, which means that the relaxation time is much larger than the timescale of interest (except perhaps at the very center). This property could be very useful: since a particle's orbit is basically what it would be if it were moving in a smooth gravitational field, we could evaluate the field instead of calculating all stellar interactions, this is cheaper computationally. Another useful property is that galaxies are often spheroidal in shape. Even highly flattened galaxies will have a spherical dark halo component. Thus, a spherical shape could be used as a zeroth order approximation for the gravitational field, and higher order terms could be written using spherical harmonics.

The goal of this paper is to examine two techniques that utilize both these facts. These are the Multipole Expansion (MEX) and the Self-Consistent Field (SCF) methods. They historically come from different ideas, and as explained below in detail, they are mathematically distinct. In the context of numerical simulations, however, they serve a similar function: to evaluate the gravitational force on all $N$ particles generated by this same collection of particles, in a way that discards spurious small scale structure (in other words, smooths the field).

MEX was born of the need to ease the computational burden. The idea is that given spherical symmetry, Gauss's law says that the gravitational force on a particle at radius $r$ from the center is simply $GM(r)/r^2$, towards the center, where $M(r)$ is the enclosed (internal) mass. The gravitational constant, $G$, will be omitted in the following text. This idea was used by \citet{Henon64} who simulated clusters with up to 100 particles to study phase mixing due to spherical collapse. This ``spherical shells'' methods is MEX of order zero and was also used for the same purpose by \citet{Bouvier+70}. The extension of this this idea is that when spherical symmetry breaks, corrections to the force can be expressed by summing higher multiples (dipole, quadruple, etc.) of the other particles, both internal and external to $r$. \citet{Aarseth67} used such a code to study a stellar cluster of a 1\,000 stars embedded in a galactic potential, truncating the expansion at $\lmax = 4$.

\citet{vanAlbada+77} used a variation of this method to study galaxy collision. These authors employed a grid and conducted simulations of also up to $N = 1\,000$ and $\lmax = 4$. They additionally assumed azimuthal symmetry which reduced the number of terms in the expansion. \citet{Fry+80}, \citet{Villumsen82}, \citet{McGlynn84} and \citet{White83} all use variations of this method, with additional features which are partly discussed in Secion \ref{sec:discussion-mex}. See also \citet{Sellwood87} for a review.

The prehistory of SCF is rooted in the problem of estimating a disk galaxy's mass distribution from its rotation curve. \citet{Toomre63} proposed a mathematical method to generate a surface density profile and a corresponding rotation curve (related to the potential) by means of a Hankel transform, and introduced a family of such pairs. \citet{Clutton-Brock72} used Toomre's idea, but in reverse: to calculate the gravitational field from an arbitrary 2D density, he generated an orthogonal set of density profiles and their corresponding potentials. This solved two problems (1) with his orthogonal set it was possible to represent any flat galaxy as a finite linear combination of basis functions, and (2) unwanted collisional relaxation was curbed due to the smooth nature of the reconstructed gravitational field. Cf. a related method by \citet{Schaefer+73}. \citet{Clutton-Brock73} introduced a 3D extension of his method, which was called SCF by \citet[hereafter HO92]{HO92} by analogy to a similar technique used in stellar physics \citet{Ostriker+68}; further historical developments are discussed in Section \ref{sec:radial-basis}.

To exploit recent developments in the world of general purpose computing on GPUs, we implemented both SCF and MEX routines in a code called \textit{ETICS} (acronym for \textit{Expansion Techniques in Collisionless Systems}). In Section \ref{sec:formalism} we explain the mathematical formalism of both methods and highlight the differences between them. In Section 3 we explain the unique challenges in a GPU implementation and measure the code's performance. In Section 4 we discuss the accuracy of expansion and direct techniques. In Section 5 we present a general discussion and finally summarize in Section 6.

\subsection{Glossary}
Here we clarify some terms used throughout this work:

\begin{description}
\item [Expansion methods] a way to get potential and force by summing a series of terms; in this paper either MEX or SCF.
\item [MEX] Multipole expansion method (sometimes known in the literature as the Spherical Harmonics method); expansion of the angular part.
\item [SCF] Self-consistent field method; a ``pure'' expansion method since both angular and radial parts are expanded.
\item[\textit{ETICS}] Expansion Techniques in Collisionless Systems; the name of the code we wrote, which can calculate the force using both MEX and SCF, using a GPU.
\item[GPU] Graphics Processing Unit; a chip with highly parallel computing capabilities, originally designed to accelerate image rendering but is also   used for general-purpose computing. It often lies on a video card\footnote{Many GPUs lie on GPU accelerator cards which lack video output.} that can be inserted into an expansion slot on a computer motherboard.
\end{description}

\section{Formalism}\label{sec:formalism}

\subsection{Series Expansions}

Both MEX and SCF methods are ways to solve the Poisson equation:
\begin{equation}
\nabla^{2}\Phi(\mathbf{r})=4\pi\rho(\mathbf{r}),\label{eq:poisson}
\end{equation}
the formal solution of which is given by the integral:
\begin{equation}
\Phi(\mathbf{r})=-\int\frac{\rho(\mathbf{r}')\mathrm{d}^{3}r'}{|{\bf r}-{\bf r}'|}.\label{eq:pot-3D}
\end{equation}
The expression $|{\bf r}-{\bf r}'|^{-1}$ is the Green's function of the Laplace operator in three dimensions and in free space (no boundary conditions), and the integral is over the whole domain of definition of $\rho(\mathbf{r})$. In an $N$-body simulation, the density field $\rho(\mathbf{r})$ is sampled at $N$ discrete points $\{\mathbf{r}_{j}\}$, such that
\begin{equation}
\rho(\mathbf{r})=\sum_{j=1}^{N}m_{j}\delta(\mathbf{r}-\mathbf{r}_{j}),
\end{equation}
where $\delta(\mathbf{r})$ is the 3D Dirac delta function. Direct $N$-body techniques evaluate integral (\ref{eq:pot-3D}) \textit{directly}:
\begin{equation}
\Phi(\mathbf{r})=-\sum_{j=1}^{N}\frac{m_{j}}{|{\bf r}-{\bf r}_{j}|},
\end{equation}
and thus at each point $\mathbf{r}=\mathbf{r}_{i}$ where the potential is evaluated, require $N$ calculations of inverse distance, or $N-1$ if $\mathbf{r}_{i}\in\{\mathbf{r}_{j}\}$, since there is no self-interaction. In practice, we are interested in evaluating the potential at the same points in which the density field is sampled, and thus a ``full'' solution of the Poisson equation requires $N(N-1)/2\approx N^{2}$ inverse distance calculations. In both MEX and SCF the integrand in equation (\ref{eq:pot-3D}) is expanded as a series of terms, each of which more easily numerically integrable; this is done in two different ways, lending the two methods quite different properties. In both methods, the reduction in numerical effort comes at the expense of accuracy compared to direct-summation, but this statement is arguable since in practice direct $N$-body techniques use a very small number of particle to sample the phase space.

To demonstrate the difference between the two approaches in the following Section, let us consider a 1D version of integral (\ref{eq:pot-3D}); let us further assume that the density exists in the interval $0\leq x\leq1$:
\begin{equation}
I(x)=\int_{0}^{1}\frac{\rho(x')\mathrm{d}x'}{|x-x'|}=\int_{0}^{x}\frac{\rho(x')\mathrm{d}x'}{x-x'}-\int_{x}^{1}\frac{\rho(x')\mathrm{d}x'}{x-x'}.\label{eq:pot-1D}
\end{equation}
Note that this is not a solution for a 1D Poisson equation (hence the notation $I$ instead of $\Phi$), but just a simplification we will use to illustrate the properties of each method. We will conveniently ignore the fact that this integral is generally divergent in 1D, as it does not affect the following discussion. In brief, MEX is a \textit{Taylor}-like expansion of the Green's function, while SCF is a \textit{Fourier}-like expansion of the density. This already hints at the most critical difference between the MEX and SCF: while the former, like a Taylor series, is local in nature, the latter is global. Another way to look at it is that in both methods the integrand is written as a series of functions (of $x$) with coefficients: in MEX one uses the given density to evaluate the functions, while their coefficients are known in advance; in SCF one evaluates coefficients, while the functions are known in advance.

\subsection{MEX}

Let us define $z\equiv x'/x$ and expand the Green's function equivalent in equation (\ref{eq:pot-1D}) around $z=0$, we get that for $z<1$ or $x'<x$:
\begin{equation}
\frac{1}{|x-x'|}=\frac{1}{x}\frac{1}{|1-z|}=\frac{1}{x}\sum_{l=0}^{\infty}z^{l},
\end{equation}
while for $z>1$ or $x'>x$ we can expand around $z^{-1}=0$:
\begin{equation}
\frac{1}{|x-x'|}=\frac{1}{x'}\frac{1}{|z^{-1}-1|}=\frac{1}{x'}\sum_{l=0}^{\infty}z^{-l}.
\end{equation}
The first and second terms of integral (\ref{eq:pot-1D}) define the functions $q_l(x)$ and $p_l(x)$ (utilizing the commutativity of the sum and integral operations):
\begin{eqnarray}
\int_{0}^{x}\frac{\rho(x')\mathrm{d}x'}{|x-x'|} & = & \sum_{l=0}^{\infty}x^{-(l+1)}\int_{0}^{x}\rho(x')x^{\prime l}\mathrm{d}x'\equiv\sum_{l=0}^{\infty}x^{-(l+1)}q_{l}(x),\\
\int_{x}^{1}\frac{\rho(x')\mathrm{d}x'}{|x-x'|} & = & \sum_{l=0}^{\infty}x^{l}\int_{x}^{1}\rho(x')x^{\prime-(l+1)}\mathrm{d}x'\equiv\sum_{l=0}^{\infty}x^{l}p_{l}(x),
\end{eqnarray}
and thus
\begin{equation}
I(x)=\sum_{l=0}^{\infty}\left[q_{l}(x)x^{-(l+1)}+p_{l}(x)x^{l}\right].
\end{equation}
While seemingly we made things worse (instead of one integral to evaluate, we now have a series of integrals), the fact that $x$ has moved from the integrand to the integral's limit greatly simplifies things. Let the density $\rho(x)$ be sampled at $N\gg1$ discrete and ordered points $\{x_{j}:x_{j}<x_{j+1}\}$; it is easy to show that
\begin{eqnarray}
q_{l}(x_{i}) & = & \sum_{j<i}m_{j}x_{j}^{l}\\
p_{l}(x_{i}) & = & \sum_{j>i}m_{j}x_{j}^{-(l+1)}
\end{eqnarray}
In other words, each of these functions is a cumulative sum of simple terms and can be evaluated at all $\{x_{j}\}$ in just one pass, but a sorting operation is required.

\subsection{SCF}

Let us leave the Green's function as it is, and instead expand the density as a generalized Fourier series:
\begin{eqnarray}
\rho(x) & = & \sum_{n=0}^{\infty}a_{n}\rho_{n}(x)\label{eq:SCF-sum}\\
a_{n} & = & \int_{0}^{1}\rho(x')\rho_{n}^{*}(x')\mathrm{d}x'\nonumber 
\end{eqnarray}
where $\{\rho_{n}(x)\}$ is a complete set of real or complex functions (the basis functions); orthonormality of the basis functions is assumed above. The integral (\ref{eq:pot-1D}) becomes:
\begin{equation}
I(x)=\sum_{n=0}^{\infty}a_{n}\int\frac{\rho_{n}(x')\mathrm{d}x'}{|x-x'|}\equiv\sum_{n=0}^{\infty}a_{n}I_{n}(x).
\end{equation}
The function set $\{I_{n}(x)\}$ is defined by the above integral. In essence, we replaced the integral over an arbitrary density $\rho(x)$ with an integral over some predefined `densities' $\rho_{n}(x)$. The advantage is that we can calculate the corresponding potentials, $I_{n}(x)$ in advance, and then the problem is reduced to numerically determining the coefficients $a_{n}$. The choice of the basis is not unique, and an efficient SCF scheme requires that the following:
\begin{enumerate}
\item The functions $\rho_{n}(x)$ and $I_{n}(x)$ are easy to evaluate numerically.
\item The sum (\ref{eq:SCF-sum}) convergence quickly, or in other words, $\rho_{0}(x)$ is already close to $\rho(x)$.
\end{enumerate}

\subsection{Properties in Three Dimensions}\label{sec:3d}

The standard form of MEX in 3D is
\begin{eqnarray}
\Phi({\bf r}) & = & -\sum_{l=0}^{\infty}\frac{4\pi}{2l+1}\sum_{m=-l}^{l}\left[q_{lm}(r)r^{-(l+1)}+p_{lm}(r)r^{l}\right]Y_{lm}(\theta,\phi)\label{eq:mex-phi}\\
q_{lm}(r) & = & \int_{r'<r}r^{\prime l}\rho({\bf r}')Y_{lm}^{*}(\theta',\phi'){\rm d}^{3}r'\label{eq:mex-qlm}\\
p_{lm}(r) & = & \int_{r<r'}r^{\prime-(l+1)}\rho({\bf r}')Y_{lm}^{*}(\theta',\phi'){\rm d}^{3}r'.\label{eq:mex-plm}
\end{eqnarray}
All together there are $\frac{1}{2}(\lmax+1)(\lmax+2)$ complex function pairs (not counting negative $m$, which are complex conjugates of the others) that need be calculated from the density. Since in practice the density field is made of $N$ discrete points, they must be sorted by $r$ in order for the above integrals to be evaluated in one pass.

The standard form for SCF is:
\begin{equation}
\Phi({\bf r})=\sum_{n=0}^{\infty}\sum_{l=0}^{\infty}\sum_{m=-l}^{l}A_{nlm}\Phi_{nl}(r)Y_{lm}(\theta,\phi).\label{eq:scf-phi}
\end{equation}
All together there are $\frac{1}{2}(\nmax+1)(\lmax+1)(\lmax+2)$ complex coefficients (not counting negative $m$) that need be calculated from the density. A typical choice is $(\nmax,\lmax)=(10,6)$, for which there would be 308 coefficients. The radial basis functions and coefficients for SCF are discussed in the next Section. Spherical harmonics are used in both cases to expand the angular part, but alternatives exist, such as spherical wavelets (e.g. \citealt{Schroder+95}). MEX has two sums (one infinite) while SCF has three sums (two infinite). In practice, the radial and angular infinite sums must be cut off at $n_{\mathrm{max}}$ and $l_{\mathrm{max}}$, respectively. The finite sum could in principle also be truncated to discard azimuthal information.

Simply equating the expressions gives the relation between the two methods:
\begin{equation}
Q_{lm}(r)\equiv-\frac{4\pi}{2l+1}\left[q_{lm}(r)r^{-(l+1)}+p_{lm}(r)r^{l}\right]=\sum_{n=0}^{\infty}A_{nlm}\Phi_{nl}(r),
\end{equation}
where $Q_{lm}(r)$ is the $(l,m)$-pole. In case the system is azimuthally symmetric, $Q_{lm}=0$ for all $|m|>0$. Also, the same azimuthal information is carried in positive and negative $m$ terms, and they are related to each other by complex conjugation.

If one decompose the density to a spherical average $\bar{\rho}(r)$ and the non-spherical deviation $\tilde{\rho}(r,\theta,\phi)$, then it is easy to show that $Q_{00}(r)$ depends only on the spherical average, while all other term depend only on the deviation. In a spherically symmetric system only $Q_{00}$ is nonzero, and setting $l_{\mathrm{max}}=0$ yields an accurate result. While the choice of $l_{\mathrm{max}}$ depends only on the deviation of the system from spherical symmetry, the choice of $n_{\mathrm{max}}$ in SCF depends on how well the system is described by the the zeroth radial basis function, and is usually determined by trial and error (see Section \ref{sec:inifinite-particle}).

It is interesting to note a nontrivial mathematical difference between the two methods. One can show that the Laplacian of equation (\ref{eq:mex-phi}) is zero when substituting the appropriate expressions for $q_{lm}$ and $p_{lm}$ from the equations (\ref{eq:mex-qlm}) and (\ref{eq:mex-plm}); the proof is mathematically cumbersome and will not be brought here. This is surprising, since according to the Poisson equation the result should be proportional to the density. One cannot appeal to series truncation to resolve this apparent contradiction; indeed each term in the formally non-truncated infinite series yields a zero density, despite representing the multipoles as continuous functions. The solution is that the potential at point $\mathbf{r}$ has contributions from all internal (i.e. at $r'<r$) particles (represented by $q_{lm}$) and all external particles (represented by $p_{lm}$), but no information about the density at $\mathbf{r}$ itself. This is the case also when the potential is constructed by a direct-summation of all gravitational point sources, so one may say MEX is similar to direct methods in this sense. In SCF, by construction, taking the Laplacian of equation (\ref{eq:scf-phi}) leads right back to the density field (equation \ref{eq:poisson}). One can thus use the coefficients $A_{nlm}$ to represent a smoothened field. One can also use MEX for this purpose, if the derivatives of $Q_{lm}(r)$ are calculated on a grid or with a spline.

\subsection{Radial Basis}\label{sec:radial-basis}

A key difference between MEX and SCF is the freedom of choice of \emph{radial} basis. There are in fact two function sets: the radial densities $\{\rho_{nl}(r)\}$ and the radial potentials $\{\Phi_{nl}(r)\}$; they are related via the Poisson equation $\nabla^2\Phi_{nl}=4\pi\rho_{nl}$ (in this case $\nabla^2$ only contains derivatives with respect to $r$). The choice of basis is not unique, and the basis functions themselves need not represent physical densities and potentials (i.e. $\rho_{nl}$ could be negative). However it is convenient to take the zeroth term ($n=l=0$) to represent some physical system, and to construct the rest of the set by some orthogonalization method, such as the Gram--Schmidt process.

The idea of \citet{Clutton-Brock73} was to use a \citet{Plummer11} model as the zeroth term and construct the next orders using the Gegenbauer (ultraspherical) polynomials and spherical harmonics (cf. \citealt{Allen+90} who developed a virtually identical method for finite stellar systems using spherical Bessel functions for the radial part). HO92 constructed a new radial basis (also using Gegenbauer polynomials) which zeroth order was a \citet{Hernquist90} model; this is the basis set we adopt in \textit{ETICS}. They argued that this basis was more well suited to study galaxies.

More basis sets followed. \citet{Syer95} used the idea of \citet{Saha93}, that the basis does not have to be biorthogonal, to construct as set which zeroth order was oblate. \citet{Zhao96} gave a radial basis set for the more general $\alpha$-model (of which both Plummer and Hernquist are special cases) and \citet{Earn96} introduced a basis for thick disks in cylindrical coordinates. \citet{Brown+98}, \citet{Weinberg99} describe numerical derivation of the radial basis set so that the lowest order matches any initial spherically-symmetric model, so called ``designer basis functions''. \citet{Rahmati09} introduced an analytical set which zeroth order is the perfect sphere of \citet{deZeeuw85}.

\section{Implementation}

\subsection{Parallelism}\label{sec:implementation-parallel}

There are several levels of task parallelism available when writing computer code. At one level, tasks are performed in parallel on different computational units (such as CPUs) but only one copy of the data exists, which is accessed by all tasks; this is called a \textit{shared memory} scheme. The tasks are called ``threads'', and they are generally managed within one ``process'' of the program. A higher level of parallelism is called \textit{distributed memory} scheme, where tasks are performed on different units (often called ``nodes''), but each unit has access only to its own memory; thus data must be copied and passed. In this case the parallel tasks are different processes, and cooperation between them is facilitated by a message passing interface (MPI). The parallel programming model is different between shared and distributed memory; the former is considered easier since threads can faster and more easily cooperate. A high-performance supercomputer will generally enable parallelism on both levels: these machines are made of multiple nodes, each of which has its own memory and multiple computational units.

Graphics processing units (GPUs) are powerful and cost-effective devices for high performance parallel computing. They are used to accelerate many scientific calculations, especially in astrophysics, such as dynamics of dense star clusters and galaxy centers (\citealt{Hamada+07}; \citealt{PZ+07}; \citealt{Schive+08}; \citealt{Just+11}; see review by \citealt{Spurzem+12}). The GPU contains its own memory and many computational units, thus it is a shared memory device\footnote{A GPU behaves as a shared memory device since all threads have transparent access to the device's global memory, which has a single address space. However there is a hierarchy in the memory and thread structure, with some kinds of memory private at the thread or block level. Thus, GPUs has also distributed memory characteristics.}. SCF force calculation is particularly easy to parallelize, since the contribution of each particle to the coefficients $A_{nlm}$ is completely independent of all other particles. Particle data can be split to smaller chunks (each could be on a different node), from each chunk partial $A_{nlm}$-s are calculated. Then the partial coefficients summed up and the result communicated to all the nodes. This was done by \cite[hereafter H95]{Hernquist+95}, whose code used the MPI call \texttt{MPI\_Allreduce} to combine the partial coefficients. This parallelization scheme, however, is not suitable for GPUs, as discussed in Section \ref{sec:implementation-scf}. MEX force calculation is harder to parallelize since the contribution of each particle depends on its position in a sorted list (by radius). However, in a shared memory scheme this too could be achieved relatively easily as explained in the following Section.

\subsection{MEX}\label{sec:implementation-mex}

\begin{figure}
\begin{center}
\includegraphics{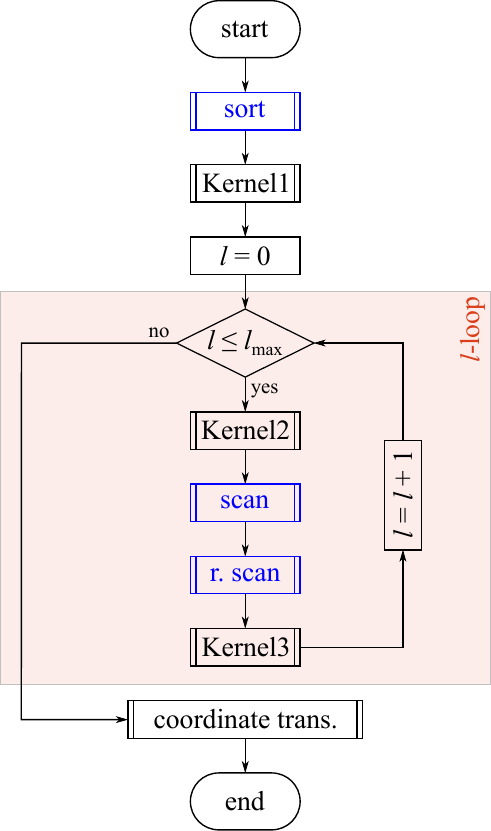}
\end{center}
\caption{Flowchart of the MEX routine. Boxes with double-struck vertical edges indicate a GPU-accelerated operation. Blue color represents call to the represent \textit{Thrust} library.}
\label{fig:flowchart-MEX}
\end{figure}

\begin{figure*}
\begin{center}
\includegraphics{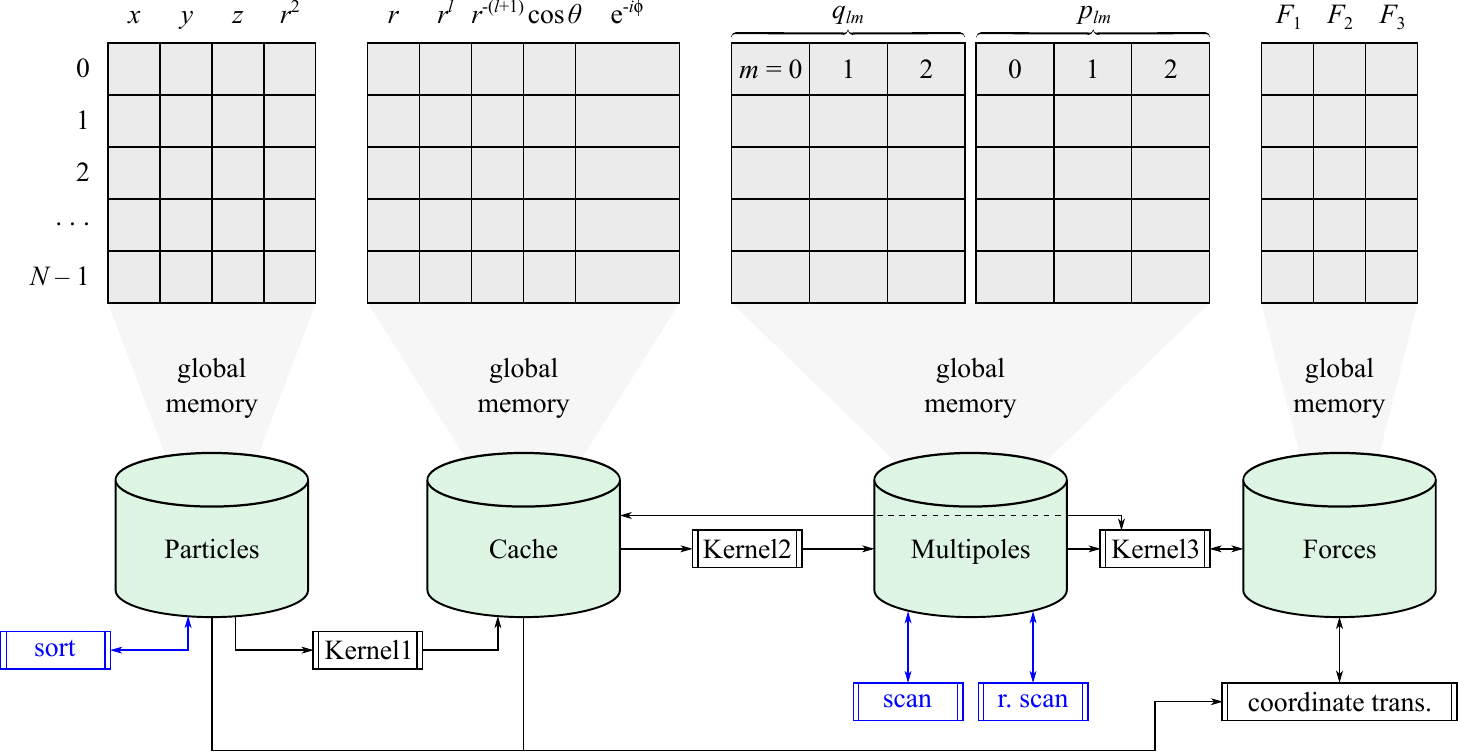}
\end{center}
\caption{The main memory structures in the MEX routine, and a scheme of the action of the \textit{Thrust} subroutines (blue) and kernels on them. The wider boxes for the exponent (in the cache structure) and the multipoles represent complex numbers (require twice the memory). The layout of the multipole structure (shown here for $\lmax=2$) is actually rotated in memory by $90^\circ$ with respect to the other structures, since it is easier for the scan subroutines.}
\label{fig:memory-MEX}
\end{figure*}

The current implementation of the MEX method relies on \textit{Thrust} \citep{Bell+11}, a C++ template library of parallel algorithms which is part of the CUDA framework. It makes parallel programming on a shared memory device (either a GPU or a multicore CPU) transparent, meaning that the task is performed in parallel with a single subroutine call, and the device setup and even choice of algorithm is performed by the library. \textit{Thrust} provides a sorting routine that selects one of two algorithms depending on input type. In the current version of MEX and using version 1.6 of \textit{Thrust}, a general Merge Sort algorithm \citep{Satish+09} is used.

A flowchart of the entire MEX routine is shown in Fig. \ref{fig:flowchart-MEX}. The flow is controlled by the CPU, and boxes with double-struck vertical edges indicate a GPU-accelerated operation. The blue double-struck boxes represent \textit{Thrust} calls, while the black ones are regular CUDA kernel calls. When a GPU operation is in progress, the CPU flow is paused. Fig. \ref{fig:memory-MEX} shows the four main memory structures of the program and how the \textit{Thrust} subroutines and kernels in the program operate on them. The particle array contains all particle coordinates and also the distance square from the center, which needs to reside in this structure for the sorting operation (in practice the particle array contains additional data such as ID and velocity, but this is not used by the MEX routine); the cache structure contains functions of particle coordinates which are needed to calculate the multipoles. Kernel1, which is executed once, reads the coordinates, calculates those functions and fills the cache structure.

Kernel2 calculates the spherical harmonics at the current $l$-level and from that the contribution of the particle to $q_{lm}$ and $p_{lm}$, which are saved in global memory. When this kernel returns, the \textit{Thrust} subroutines are dispatched to perform the cumulative sum. The ``scan'' (forward cumulative sum) and ``r. scan'' (reverse scan) are both in fact calls to the \texttt{exclusive\_scan} subroutine, but to perform the reverse scan, we wrap $p_{lm}$ with a special \textit{Thrust} structure called \texttt{reverse\_iterator}. Not shown in the flowchart, the two scan subroutines have to be called $l$ times at each $l$-level since they work on one $m$ value at a time.

Kernel3 has both cache and compute operations: it calculates the partial forces in spherical coordinates (i.e. the $l$-order correction to the force) and/or potentials by evaluating all the spherical harmonics again (and their derivatives with respect to spherical coordinates). Later it advances $r^l$ and $r^{-(l+1)}$ to the next $l$-level (except at the last iteration). Finally, the last kernel operates on the force structure and transforms it to Cartesian coordinates. Fig. \ref{fig:pie-mex} shows the relative time it takes to do the internal operation.

We note that the potential could be calculated at the same time as the force (in Kernel3) and stored at another memory structure (not shown in Fig. \ref{fig:memory-MEX}) but is skipped if only forces are needed. Alternatively, only the potential could be calculated (this is faster since the derivatives of the special functions are not calculated). The choice between calculating force, potential or both is done with C++ templates.

\subsection{SCF}\label{sec:implementation-scf}

\newcommand{\Anlm}{$A_{nlm}$}

We first briefly explain the serial algorithm used by HO92. The force (and potential) calculation had two parts: (1) calculation of all the \Anlm-s (the plural suffix `-s' to emphasize that there are hundreds of coefficients in this 3D structure) and (2) calculation of all the forces using the coefficients.

In both parts, the particle loop (the $j$-loop) was the external one, inside of which there are again two main steps. In step (1a) all necessary special functions were calculated using recursion relations. Step (2a) was identical but additionally, the derivatives of those functions were calculated. In step (1b) there was a nested loop ($n$-$l$-$m$ structure) in which a particle's contribution to every \Anlm was calculated and added serially. In step (2b) there was also such a loop, which used all the \Anlm-s to calculate the force on each particle.

In the parallel algorithm used by H95, another part was added between the two parts mentioned above: communicating all partial \Anlm-s from the various MPI processes, adding them up and distributing the results. In practice it was achieved using just one command, \texttt{MPI\_Allreduce}. There are two main reasons why this algorithm could not be used effectively on a GPU, both are related to the difference between how the GPU and CPU access and cache memory. The first difficulty is performing the sum. The partial sums from the different parallel threads could in principle be stored on a part of the GPU memory called \textit{global memory}, and then summed in parallel. However a modern GPU can execute tens of thousands of threads per kernel (note that the concept of a thread in CUDA is abstract, and the number of threads by far exceed the number of thread processors on the GPU chip), and every partial \Anlm\ is $~5$ kilobyte in size (depending on $\nmax$ and $\lmax$). Thus, writing and summing the partial coefficients would require extensive access to global memory, which is slow compared to the actual calculation part. The second difficulty is that if one thread uses too much memory, for example to store all necessary Legendre and Gegenbauer polynomials as well as complex exponent (as is done in the HO92 code), this may lead to an issue called \textit{register spilling}, where instead of using the very fast register memory, the thread will store the values on the slow global memory, which again we wish to avoid on performance grounds.

To tackle those issues we utilized another type of GPU memory called \textit{shared memory}\footnote{Not to be confused with the concept of a shared memory device.}. This memory is ``on chip'' (on the multiprocessor circuit rather than elsewhere on the video card) and has $\times 100$ lower latency than global memory. Threads in a CUDA program are grouped into blocks, threads in the same block share this fast memory (hence the name). It is also much less abundant than global memory. The Nvidia Tesla K20 GPUs have just 64 kilobytes of shared memory per block, while they have 5 gigabytes of global memory.

In order to use shared memory to calculate the coefficients, each thread would serially add contributions from particles to the partial \Anlm-s on shared memory; then they would be summed up in parallel in each block. However, there are usually hundreds of different \Anlm-s, as well as tens or hundreds of threads per block (depending on hardware; which is required for efficient loading of the GPU); there is not enough shared memory for that (by far). To solve this, we changed the order of the loops: the external loop is the $l$-loop, then comes the $n$-loop. For each $(n,l)$ pair, a CUDA kernel is executed where the $j$-loop is performed in parallel on different threads, inside of which the $m$-loop is done. Now each threads has to deal with far fewer \Anlm-s (no more than $\lmax+1$), for which there is usually enough shared memory.

\begin{figure}
\begin{center}
\includegraphics{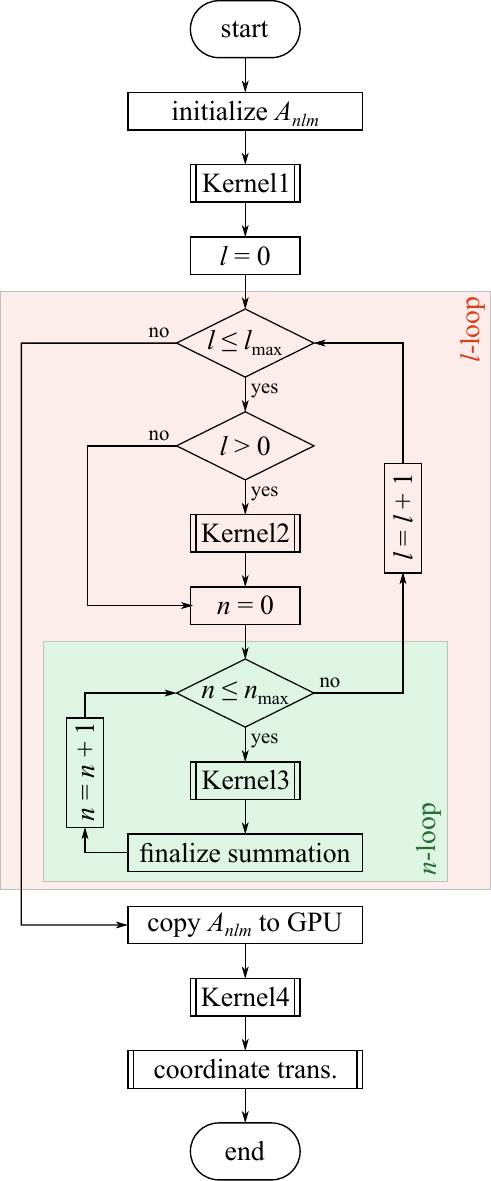}
\end{center}
\caption{Same as Fig. \ref{fig:flowchart-SCF} but for the SCF routine.}
\label{fig:flowchart-SCF}
\end{figure}

\begin{figure*}
\begin{center}
\includegraphics{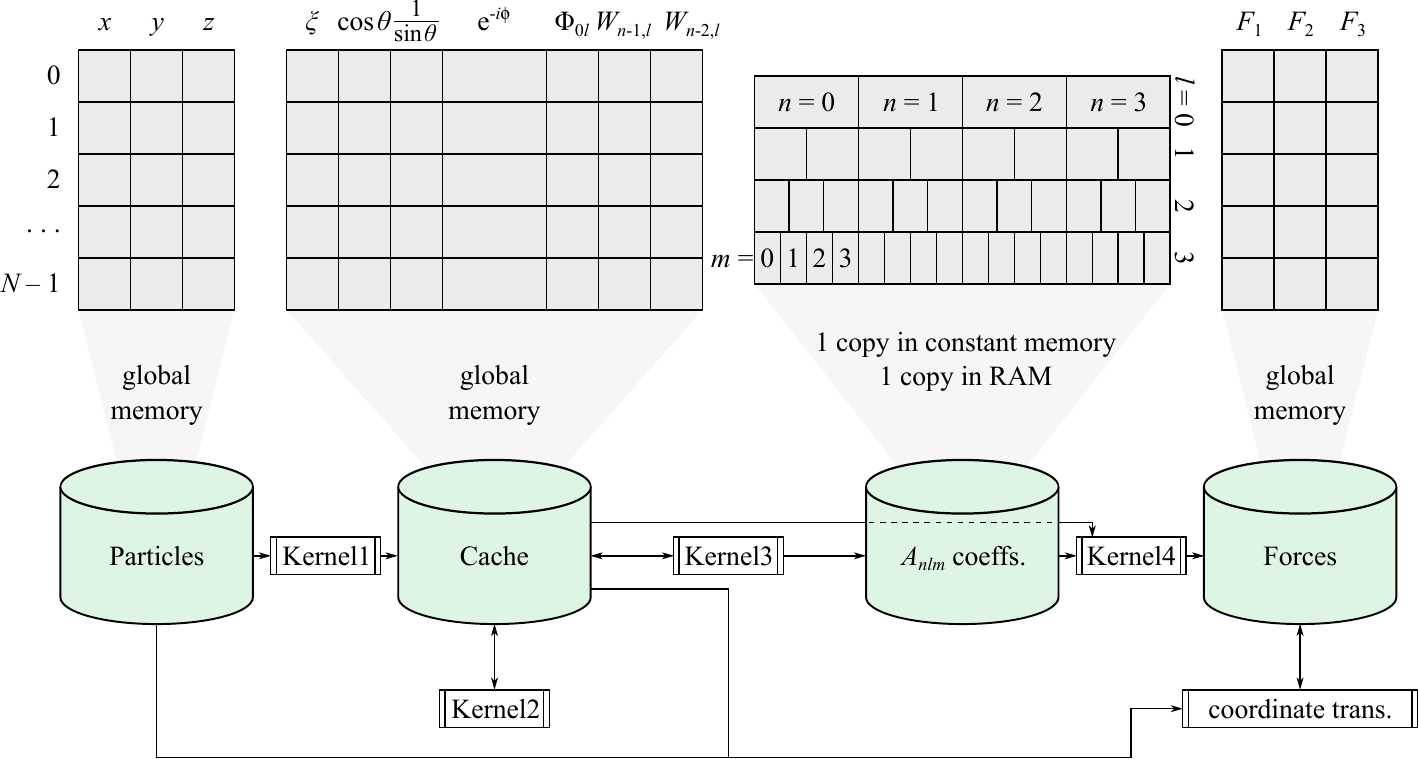}
\end{center}
\caption{Same as Fig. \ref{fig:memory-MEX} but for the SCF routine. The main memory structures in the SCF routine, and a scheme of the action of the kernels on them. Every cell in the coefficient structure (shown here for $(\nmax,\lmax)=(3,3)$) is a complex number.}
\label{fig:memory-SCF}
\end{figure*}

A flowchart of the entire SCF routine is shown in Fig. \ref{fig:flowchart-SCF}. The flow is controlled by the CPU, and boxes with double-struck vertical edges indicate a CUDA kernel call. When a GPU operation is in progress, the CPU flow is paused. Fig. \ref{fig:memory-SCF} shows the four main memory structures of the program and how the five kernels in the program operate on them. The particle array contains all particle coordinates (in practice it contains additional data such as ID and velocity, but this is not used by the SCF routine); the cache structure contains functions of particle coordinates which are needed to calculate the basis functions. Kernel1, which is executed once, reads the coordinates, calculates those functions and fills the cache structure. Kernel2 only operates on the cache structure, it has just one function which is to advance $\Phi_{0l}$ by one level; thus it needs to be executed at the beginning of each iteration of the $l$-loop. As shown in the flowchart, it is skipped for $l=0$ because Kernel1 calculates and caches $\Phi_{00}$.

Kernel3 has both cache and compute operations: it calculates the current $W_{nl}$ using recursion relations from the cached $W_{n-1,l}$ and $W_{n-2,l}$ and then updates the cache. Later it calculates the spherical harmonics and from that the contribution of the particle to the all \Anlm in the current ($n$,$l$)-level, which are saved in shared memory. When all threads in the block have finished calculating contributions of the particles assigned to them, they are synchronized and a parallel reduction is performed. Since threads from different blocks cannot share memory, the data from each block must be transfered to the host machine's memory and the CPU finishes the summation process. 

For the force calculation, just a reading the \Anlm-s is required. The GPU has yet another type of memory which is ideal for storing of coefficient or constant parameters. It is fittingly called ``constant memory'', and is as fast as shared memory when every thread in a warp accesses the same memory element. It is also very limited (usually to 64 kilobytes per device), but the \Anlm structure could still fit there nicely. Once calculation of all the coefficients is complete, it is transferred back to the GPU constant memory to be used to calculate the forces. Since only reading the coefficient is required, in Kernel4 which calculates the forces and/or potentials by evaluating all the basis functions again (and their derivatives with respect to spherical coordinates), the $j$-loop is the external one. To avoid register spilling we keep the internal loop structure as $l$-$n$-$m$, and thus we only need to recalculate the complex exponents, which is relatively cheap. Finally, the last kernel operates on the force structure and transforms it to Cartesian coordinates. Fig. \ref{fig:pie-scf} shows the relative time it takes to do the internal operation.

We note that the potential could be calculated at the same time as the force (in Kernel4) and stored at another memory structure (not shown in Fig. \ref{fig:memory-SCF}) but is skipped if only forces are needed. Alternatively, only the potential could be calculated (this is faster since the derivatives of the special functions are not calculated). The choice between calculating force, potential or both is done with C++ templates.

\subsection{Performance}

\begin{figure*}
\includegraphics[width=1\textwidth]{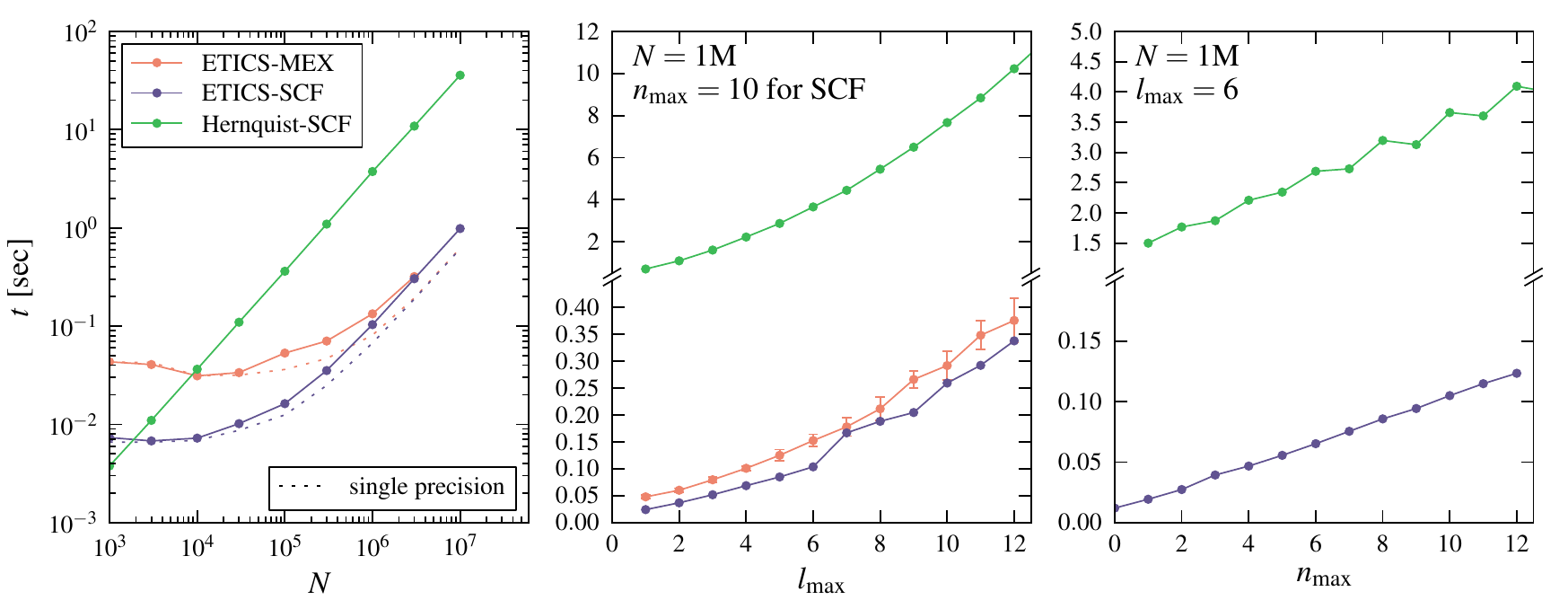}
\caption{Scaling of one full force calculation time. Hernquist's SCF code (in green) is a CPU code and was tested on Intel Xeon E5520 CPU (one core). \textit{ETICS} is a GPU code with both MEX (red) and SCF (blue) methods, and was tested with Nvidia Tesla K20 GPU; for the GPU codes, dotted lines show the performance in single-precision mode. For the scaling with $N$ we set $\lmax=6$, and for the SCF codes also $\nmax=10$. The scaling is theoretically linear with $N$ for SCF and $N\log N$ for MEX, but the theoretical behavior is only seen asymptotically for the GPU codes, since the GPU is not fully loaded at low $N$. Both methods scale quadratically with $\lmax$ (the tests were performed with $N=10^6$, and $\nmax=10$ for SCF). SCF scales linearly with $\nmax$ (the tests were performed with $N=10^6$ and $\lmax=6$). The CPU code shows some erratic behavior due to compiler optimization. Note that the tests are performed on different hardware.}
\label{fig:scaling}
\end{figure*}

We tested the performance of \textit{ETICS} (both MEX and SCF) on a single Nvidia Tesla K20 GPUs on the Laohu supercomputer at the NAOC in Beijing. For comparison, we also tested the Fortran CPU SCF code by Lars Hernquist on the ACCRE cluster at Vanderbilt University in Nashville, Tennessee (we used a node with Intel Xeon E5520 CPU). If the initial conditions are not sorted by $r$ in advance, the first MEX force calculation is more costly than all the following, since the sorting of an already nearly-sorted particle list is faster. Thus, all measurements of the MEX code are done after the system is evolved one very short leapfrog time step. Fig. \ref{fig:scaling} shows the time it takes to do one full force calculation as a function of $N$, $\lmax$ and $\nmax$. Each point represents the mean time of 10 different calculations. The dispersion is generally very low, with the exception of \textit{ETICS}-MEX with $\lmax \gtrsim 6$; only for which we show error bars. Note that the timing only depends on the number of particles (and expansion cutoffs) and not on their spatial distribution.

The CPU and GPU SCF codes are both theoretically $O(\lmax^2 \nmax N)$. At low $N$ the GPU is not fully loaded, and \textit{ETICS} performance seems superlinear with $N$. \textit{ETICS}-MEX is theoretically $O(\lmax^2 N\log N)$, but this again is an asymptotic behavior which is not observed. The lack of good GPU load for $N \lesssim 10^6$ is much more evident than the $N\log N$ nature of the algorithm. The GPU global memory was the limiting factor in how many particles could be used with both methods. The dotted lines show the performance of \textit{ETICS} using single-precision instead of double. The speed increase is 61\% for SCF and 65\% for MEX, but there is a price to pay in accuracy as noted in Section \ref{sec:single}. The speedup factor could be very different for different GPUs.

All codes should scale quadratically with $\lmax$, but as the middle panel of Fig. \ref{fig:scaling} shows, this behavior is not so clear for \textit{ETICS}-MEX. This is due to the extensive memory access this code requires, which rivals the calculation time. Memory latency on GPUs is not easy to predict; due to caching and the way memory is copied in blocks, and the latency depends not only on the amount of memory accessed but also on the memory access pattern.

SCF codes theoretically scale linearly with $\nmax$. A strange behavior of the CPU code is noted: it seems that the time increases with $\nmax$ in a ``zigzag'' fashion (the measurement error of the times is much smaller than this effect, and it is reproducible). This is paradoxical: it takes a shorter time to calculate with $\nmax=9$ than with $\nmax=8$, even though more operations are required. It is not simple to understand why this is, but it seems that the compiler performs some optimization on the first $j$-loop (coefficient computation) that only help when $\nmax$ is odd but not when it is even.

The comparison between \textit{ETICS}-GPU and Hernquist's code is not exactly fair since they use different types of hardware. Specifically for hardware we tested, \textit{ETICS}-GPU outperforms Hernquist's code by a factor of about 20 (which depends little on all parameters). However, Hernquist's code can utilize a multicore CPU (using MPI). The Xeon CPU we used has 4 cores, and two such CPUs are mounted on a single ACCRE node. We could use the Fortran code in MPI mode on all 8 effective cores with almost no overhead, and the calculation is accelerated by a factor of 8. Also, Hernquist's code calculate the jerk (force derivative), which \textit{ETICS}-GPU does not; this takes $\sim 14$ percent of the total time.

\begin{figure}
\includegraphics[width=1\columnwidth]{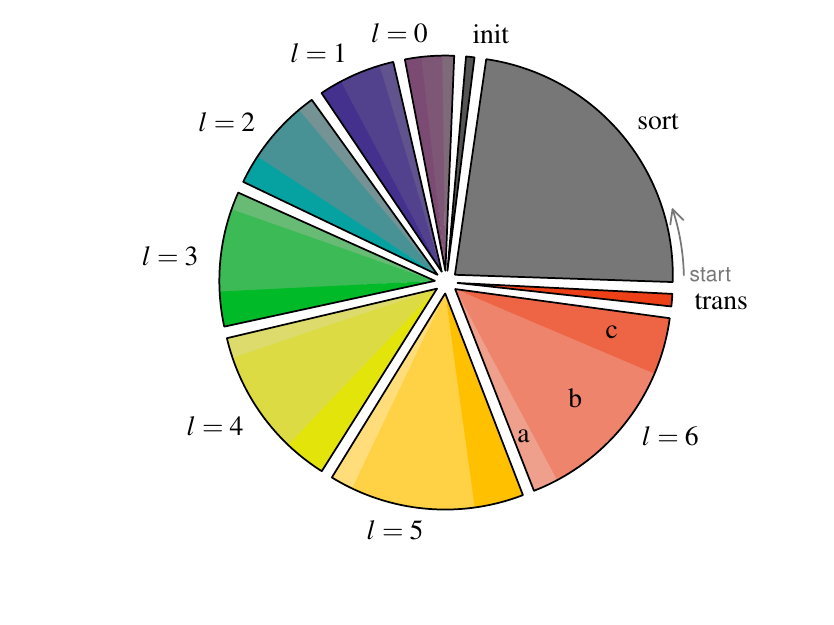}
\caption{Pie chart showing the relative time of each operation required to perform one full MEX force calculation with \textit{ETICS} (double-precision, $N=10^6$ and $\lmax=6$); the total time is 0.15 sec on Nvidia Tesla K20 GPU. The results may differ significantly on different hardware and if single-precision is used instead. The first operation is sort, followed counter-clockwise by initialization of the cache arrays, the $l$-loop where each iteration is divided to (a) summand calculation, (b) cumulative sum and (c) partial force calculation. The final operation is coordinate transformation from spherical to Cartesian}
\label{fig:pie-mex}
\end{figure}

\begin{figure}
\includegraphics[width=1\columnwidth]{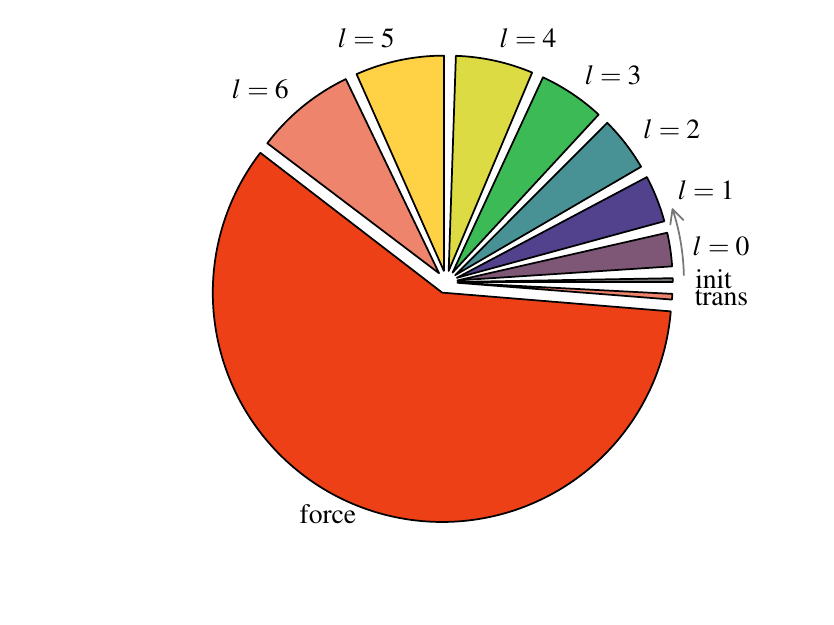}
\caption{Same as Fig. \ref{fig:pie-mex} for a full SCF force calculation with \textit{ETICS} ($N=10^6$ particles, $\lmax=6$ and $\nmax=10$); the total time is 0.16 sec on Nvidia Tesla K20 GPU. The first operation is initialization, followed counter-clockwise by the $l$-loop (in which the $n$-loop is nested). The partial force calculation is a single CUDA kernel, inside of which all the loops are performed.}
\label{fig:pie-scf}
\end{figure}

Figs. \ref{fig:pie-mex} and \ref{fig:pie-scf} show the fraction of time it takes to perform the internal operations for the force calculation for \textit{ETICS}-MEX and -SCF, respectively, both use $N=10^6$, $\lmax=6$ and for SCF, $\nmax=10$. For MEX, operations inside each iteration of the $l$-loop are shown in different shades (also denoted by letters corresponding to stages 3a, 3b and 3c as explained in Section \ref{sec:implementation-mex}). The most costly operations are the ones we entrust to \textit{Thrust}, namely the sorting and cumulative sum. In Fig. \ref{fig:pie-scf} the internal structure of each $l$-iteration is not shown (since there are too many internal operations, including the $n$-loop). The force calculation is executed as one operation (a single CUDA kernel call), and includes the $l$-loop nested inside it (unlike MEX where only a partial force was calculated at every $l$-iteration, step 3c).

\section{Expansion Accuracy}

\subsection{Infinite Particle Limit}\label{sec:inifinite-particle}

Two separate questions come up when discussing the accuracy of expansion methods: how well the expansion approximates the $N$-body force (i.e. direct-summation), and how well it approximates the smooth force in the limit of infinite particles (which we will refer to as the ``real'' force in the following discussion). Both questions depend on $N$, $\lmax$ and (for SCF) $\nmax$. A related question is how well the $N$-body force approximates the real force, as a function of $N$. All these questions depend not only on the expansion cutoff and $N$, but on the stellar distribution as well (e.g. global shape, central concentration, fractality, etc.); this will not be fully explored in this work.

There are two types of error when considering the expansion methods versus the real force, analogous to systematic and random errors. The first, systematic-like error, comes from the expansion cutoff, this is called the \textit{bias}. For example, a system which is highly flattened could not be described by keeping just the quadrupole moment, so both MEX and SCF cut off at $\lmax = 2$ would exhibit this type of error, regardless of $N$ (see \citealt{Merritt96}; \citealt{Athanassoula+00} for discussion about bias due to softening). The second, random-like error, comes from the finite number of particle and their coarse grainy distribution; it is the equivalent of $N$-body noise (also referred to as particle noise or sampling noise).

HO92 attempted to estimate accuracy of SCF by showing convergence of the coefficient amplitudes with increasing $n$ for the density profiles of some well known stellar models. They showed that $A_{n00}$ decayed exponentially or like a power law with $n$, depending on the model. This analysis was not satisfactory because it applied to the limit of infinite $N$, thus ignoring the random-like error. Furthermore, showing convergence of the coefficients does not give information about the force error. The bias and the random error are not easy to distinguish. The bias could be calculated, in principle, only if the true mass density $\rho(\mathbf{r})$ is known, which is not generally the case; however, it is still useful to look at some particular examples where it is known.

To test the accuracy of the expansions techniques, we used two simple models for the mass density. Both our models are Ferrers ellipsoids \citep{Ferrers77} (often called Ferrers bars) with index\footnote{The Ferrers index should not be confused with the SCF radial index, both denoted with $n$.} $n=1$: \texttt{model1} is a mildly oblate spheroid with axis ratio of 10:10:9, \texttt{model2} is triaxial with axis ratio of 3:2:1. Ferrers ellipsoids are often used in stellar dynamics, especially in the modeling of bars (e.g. \citealt{Athanassoula92}). They have a very simple mass density:
\begin{equation}
 \rho(\mu^{2}) = \begin{cases}
\rho_{0}\left[1-\left(\mu/a\right)^{2}\right]^{n} & \mu\leq a\\
0 & \mu>a
\end{cases}
\end{equation}
where $a$, $b$ and $c$ are the axes, $\rho_0$ is the central density, $n \geq 0$ is the index and $\mu$ is the ellipsoidal radius, defined by:
\begin{equation}
\mu^2 = x^2 + \left(\textstyle{\frac{a}{b}}y\right)^2 + \left(\textstyle{\frac{a}{c}}z\right)^2.
\end{equation}
The potential due to this family of models is simply a polynomial in $(x,y,z)$ if $n$ is an integer. The coefficients could be calculated numerically (also analytically for some cases) by solving a 1D integral \citep[Chapter 2.5]{BinneyTremaine}. For both our models we used the mathematical software \textit{Sage} to calculate the coefficients to better than $10^{-13}$. The force vector components are trivially derived from the potential polynomial; this is the ``real'' force.

\begin{figure*}
\includegraphics[width=1\textwidth]{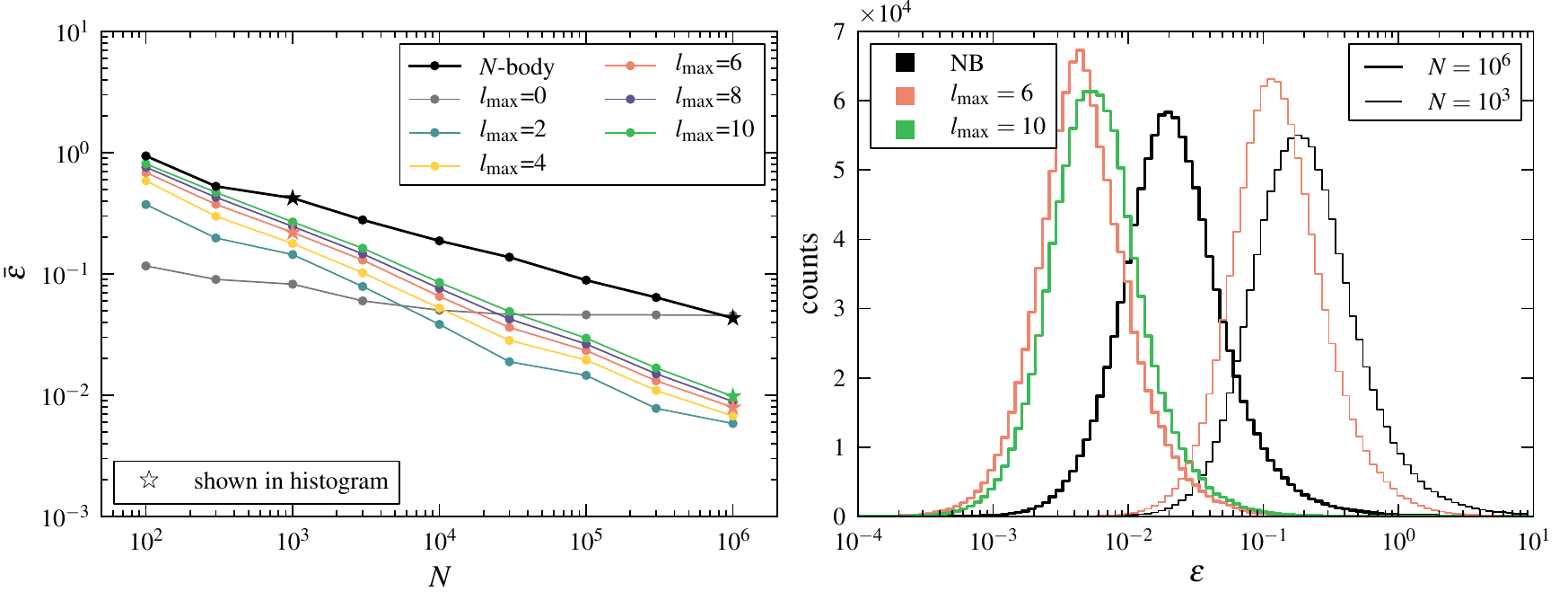}
\caption{For the mildly oblate \texttt{model1}, the left panel shows, as a function of $N$, the mean relative force error $\epsilon$ (defined in equation \ref{eq:epsilon}) in direct-summation (``$N$-body''; thick black line) as well as MEX expansions with even $\lmax$ between $0$ and $10$ (odd terms have almost no effect). Each point represents a full distribution of error values, obtain by stacking models with the same $N$. The right panel shows the full log-distribution for some selected points on the left panel (shown as stars). Notice that since the horizontal axis is logarithmic, the distributions on the right are much wider than those on the left (have larger standard deviation).}
\label{fig:force-error1}
\end{figure*}

\begin{figure*}
\includegraphics[width=1\textwidth]{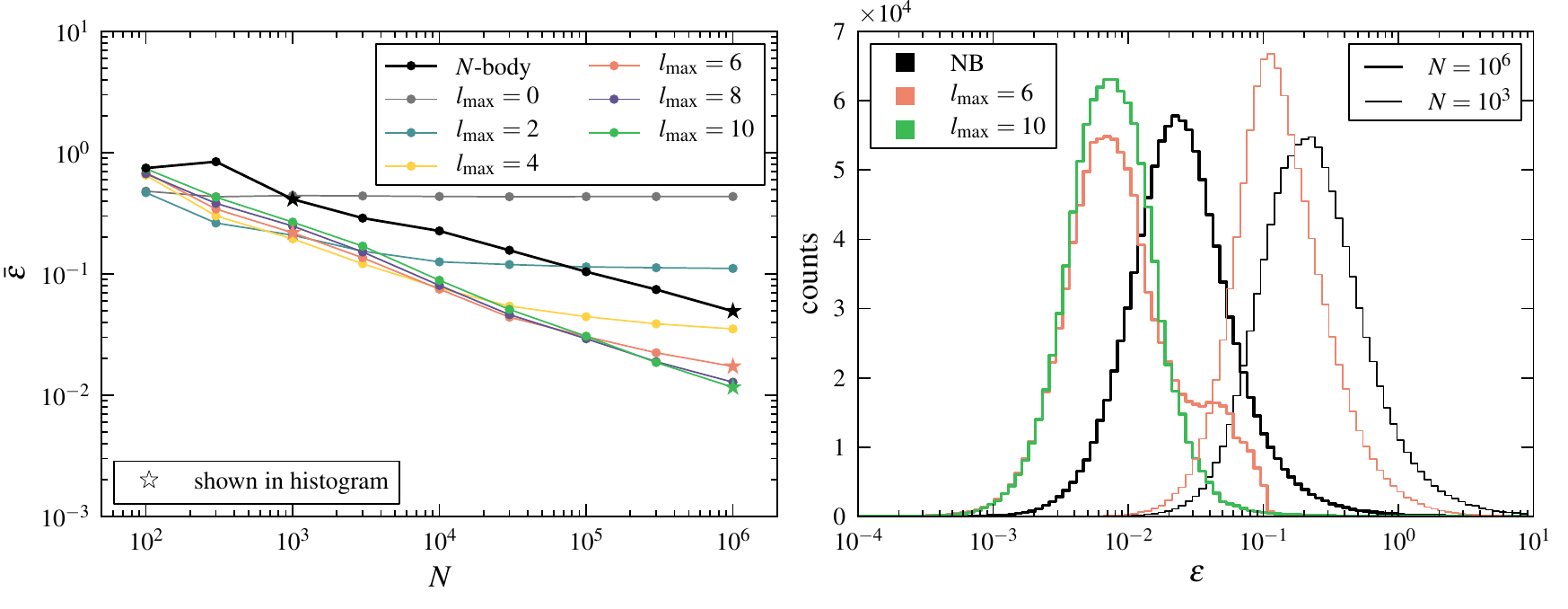}
\caption{Same as Fig. \ref{fig:force-error1} but for the triaxial \texttt{model2}. For this model, the behavior of the mean error seems different for the MEX (but nearly identical for the $N$-body): the low-multipoles outperform (have smaller mean error than) the higher ones only at low-$N$, but get saturated (increasing $N$ does not improve accuracy). This could be understood from the right panel: while the error distributions are very similar to the previous model, the $\lmax=6$, $N=10^6$ has a bump on the right. This bump is the bias that causes the saturation, and will likely not disappear when $N$ increases (however as seen it is diminished for $\lmax=10$).}
\label{fig:force-error2}
\end{figure*}

We created many realization of these two models, ranging from just 100 particles to $10^6$. The goal is to compare for each realization the force calculated using MEX, SCF and direct-summation (no softening), with the real force. All calculations performed using double-precision, and the direct-summation force is not softened. For each realization we get a distribution of $N$ values of the relative force error,
\begin{equation}
\epsilon_i \equiv \frac{|\mathbf{F}_i - \mathbf{F}_{i,\mathrm{real}}|}{|\mathbf{F}_{i,\mathrm{real}}|}\label{eq:epsilon}
\end{equation}
where $i$ is the particle's index. It is not practical to show to full distribution for all cases, so in Figs. \ref{fig:force-error1} and \ref{fig:force-error2} we show the mean, and the full distribution for only selected cases.

The left panel of Fig. \ref{fig:force-error1} shows the mean relative force error $\bar\epsilon$ in \texttt{model1} for direct-summation and MEX with even $\lmax$ between $0$ and $10$; odd terms are in principle zero if the expansion center coincides with the center of mass, and in practice very small. For this model, $\bar\epsilon$ is decreasing with $N$ for all cases but $\lmax=0$ (monopole only). The smallest error is for $\lmax=2$ (monopole and quadrupole only). Unintuitively, adding correction terms \textit{increases} the error (for constant $N$), This is because the model's deviation from sphericity is so mild, that the quadruple describes it well enough; the following terms just capture some of the $N$-body noise in the realization and make more harm than good. In the right panel we show the full log-distribution for selected cases. The histograms for $N=10^3$ are made by stacking of $10^3$ realizations, so there are $10^6$ values of $\epsilon$ in all histograms. In all cases the $\epsilon$ distributions are close to log-normal; the logarithmic horizontal axis hides the fact that the distributions on the right are much wider in terms of standard deviation due to a very long and fat tail when viewed in linear space. Note that while the number of particles increased by 1\,000, in all cases the error distribution shifted down by just a factor of $\sim 10$.

Fig. \ref{fig:force-error2} is the same but for the triaxial \texttt{model2}. While in the $N$-body cases the $\epsilon$ distributions are much the same, MEX shows a different behavior. The most prominent feature is the bump on right side of the $N=10^6$, $\lmax=6$ error distribution, which demonstrates the issue of bias. Most of the particles which make up this bump are located in the lobes of the ellipsoid, where many angular terms are required. When $\lmax$ is increased to $10$, this bump disappears. It also is not present in the $N=10^3$, $\lmax=6$ case, probably because it is overwhelmed by the random error. This bump causes the mean error to saturate with particle number, as the left panel shows. Increasing $N$ will shift the bulk of the bell curve to the left (zero), but will not quench the bump. At much larger $N$, \texttt{model1} will show the same behavior as the random error becomes smaller than the bias, and the high-$\lmax$ cases would outperform $\lmax=2$.

\begin{figure}
\includegraphics[width=1\columnwidth]{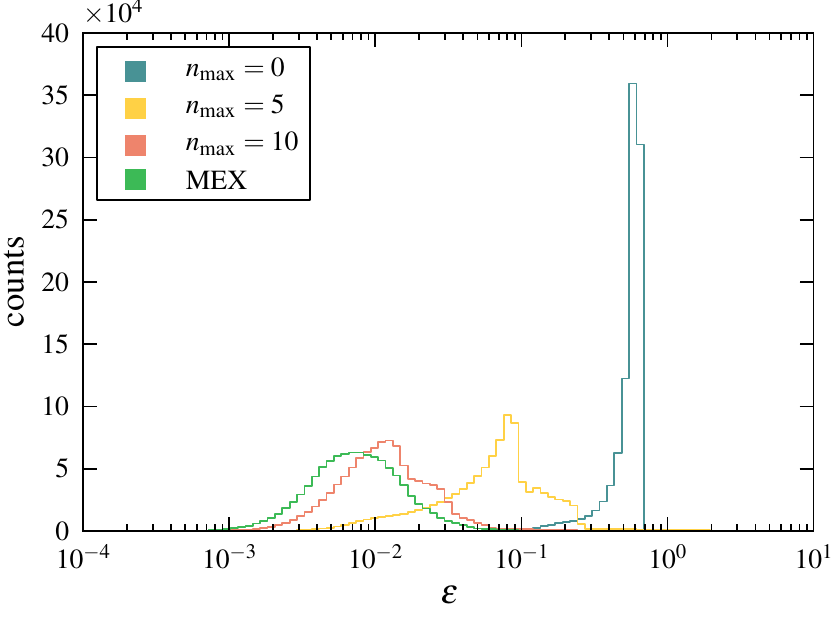}
\caption{The relative force error distribution for the triaxial \texttt{model2} with $N=10^6$. The green histogram is also shown in the right panel of Fig. \ref{fig:force-error2} and represent a MEX expansion with $\lmax=10$. The other histograms represent SCF expansions with $\lmax=10$ and varying $\nmax$ values as shown. With increasing $\nmax$, the SCF error distribution approaches that of MEX with the same $\lmax$. In this case, the model differs greatly from the zeroth order function of the basis set, showing relatively slow convergence.}
\label{fig:force-scf-error2}
\end{figure}

We repeat this exercise for SCF, which has an additional source of bias due to the radial expansion cutoff. Fig. \ref{fig:force-scf-error2} illustrates that point by showing the relative force error distribution in \texttt{model2} for SCF compared to MEX with the same number of particles ($N=10^6$) and same angular cutoff ($\lmax=10$). With increasing $\nmax$, the SCF error distribution approaches that of MEX, demonstrating the point made in Section \ref{sec:3d} that MEX is equivalent to SCF with $\nmax\rightarrow\infty$. It must be noted that the basis set we programmed in \textit{ETICS} is not at all suitable for Ferrers models (which are finite and have a flat core), and the apparently slow convergence should not disparage one from using SCF, even if it is not known in advance what basis to choose. The overlap between the relative force error distributions of SCF at $\nmax=10$ and MEX is 77\%. A more intelligent choice of basis function is discussed by \citet{Kalapotharakos+08}, who used a similar methodology to choose the best basis set for triaxial \citet{Dehnen93} $\gamma$-models \citep{Merritt+96} with $0\leq \gamma\leq1$ from a family of basis sets similar to the HO92.

The results presented in this section suggest that there is some optimal expansion cutoff, which is different for different models and depends on the number of particles \citep{Weinberg96}. This is analogous to optimal softening in direct-summation force calculations \citep{Merritt96,Athanassoula+98}. If not enough terms are used, there is a large bias; if too many terms are used, the particle noise dominates. \citet{Vasiliev13} addressed this issue by calculating the variance of each SCF coefficient among several realizations of the same triaxial Dehnen model, found that for $N=10^6$ particles, angular terms beyond $l=8$ are dominated by noise (and that only the first few $n$, $m$ terms at that l-level are reliable). 

The force error discussed above is not directly related to energy diffusion or relaxation, which are reduced due to the smoothing, but not absent. The mechanism for energy (and angular momentum) diffusion in both expansion methods is temporal fluctuation of the multipoles or coefficients (due to the particle noise). This is somewhat analogous to two-body relaxation in that the potential felt by every particle fluctuates (although in this case there is no spatial graininess). \citet[in prep.]{Vasiliev14} examined energy diffusion in a Plummer sphere with $N=10^5$ particles using SCF and direct $N$-body codes, and found that SCF demonstrated a diffusion rate only several times lower, which was close to the rate in a direct technique using near-optimal softening for this $N$. Further reduction was achieved by discarding of expansion terms which are nominally zero in any triaxial system centered around the expansion center. Finally, Vasiliev used temporal softening (HO92), where the coefficients (and thus the potential) are updated in longer intervals than the dynamical time step; this procedure however introduces a global energy errors unless some measures are taken to amend this.

\subsection{Single Precision}\label{sec:single}

Due to their original intended use, GPUs are not optimized for double-precision arithmetic (indeed early GPUs completely lacked a double-precision floating-point type). In cards that do support double-precision, arithmetic operations could still be significantly slower than for single. As noted before, in our test we measured a 60--65\% speed increase when using single-precision. The Nvidia Tesla K20 GPUs we used have enhanced double-precision performance with respect to other GPUs, for which using double-precision may be significantly slower. Those devices are somewhat specialized for scientific use and are thus more expensive (albeit in many applications still superior to parallel CPU architectures in terms of price/performance ratio due to the low energy consumption). CPUs usually take the same time to perform an arithmetic operation in either single- or double-precision, but a program's general performance could be faster in single-precision due to smaller memory load. For the Hernquist-SCF CPU code, we measured a 6\% improvement in speed. Using single-precision however inevitably reduces the accuracy of the calculated force; here we examine how bad this \textit{performance-accuracy trade-off} is.

\begin{figure}
\includegraphics[width=1\columnwidth]{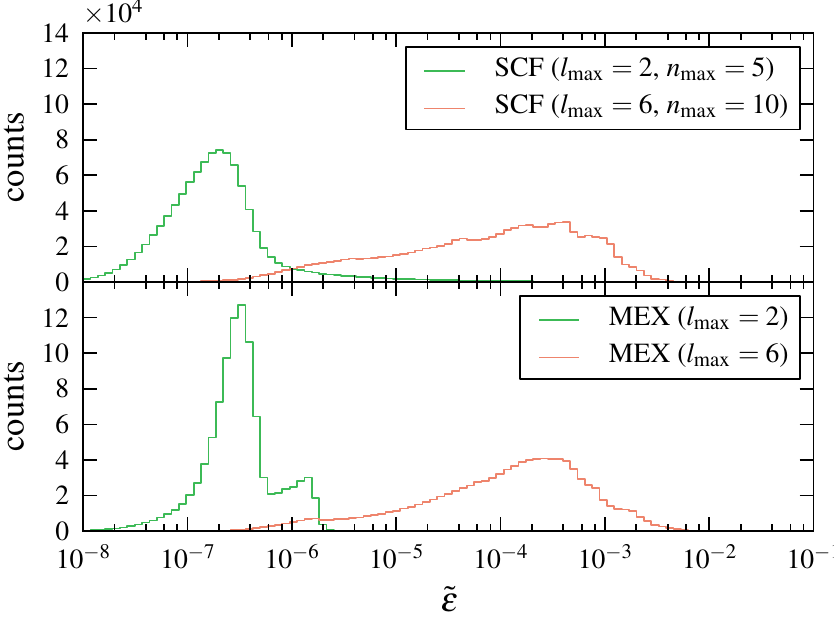}
\caption{Relative force error distributions of single-precision force expansions compared to double-precision (defined in equation \ref{eq:epsilon-single}). The model used is a Hernquist sphere with $N=10^6$. The top and bottom panels shows SCF and MEX force calculations, respectively. In both cases the left (green) histogram is a lower-order expansion as indicated in the legend.}
\label{fig:single}
\end{figure}

Fig. \ref{fig:single} show the relative force error distributions of single-precision calculations, compared to double. The relative force error on particle $i$ is now defined as:
\begin{equation}
\tilde{\epsilon}_i \equiv \frac{|\mathbf{F}_{i,\mathrm{single}} - \mathbf{F}_{i,\mathrm{double}}|}{|\mathbf{F}_{i,\mathrm{double}}|}\label{eq:epsilon-single}
\end{equation}
We testes an $N=10^6$ realization of a Hernquist sphere with characteristic scale of one unit. The top panel shows two SCF force calculations: the green histogram (on the left) is a low order expansion up to $(\nmax,\lmax)=(5,2)$, retaining 36 coefficients; the red histogram is an expansion up to $(\nmax,\lmax)=(10,2)$, retaining 308 coefficients. The bottom panel similarly shows two MEX expansions. In both cases, the higher order expansion has relatively large errors. While it is still smaller than the error with respect to the ``real'' force discussed in Section \ref{sec:inifinite-particle}, its nature is numeric and it could hinder energy conservation.

The relatively large error is not remedied by usual methods to improve accuracy of floating point arithmetic such as Kahan summation algorithm \citep{Kahan65}, because the error does not come from accumulation of round off errors. Instead, the accuracy bottle neck is the calculation of the spherical harmonics and/or the Gegenbauer polynomials. Particles for which those special functions are calculated with large numerical error
will have a large force error, but additionally they contribute erroneously to \textit{all} the coefficient or multipoles, thus causing some error in the force calculation of all other particles as well.

There are two groups of particles with large relative error in this implementation: particles that are very far away from the center, and particles which happen to lie very close to the $z$-axis. The former group is not so problematic since the absolute force is very small as well as their contribution to the coefficients or multipoles. The latter group causes large error because the recursion relation used to calculate the associated Legendre polynomials:
\begin{align}
P_{lm}(\theta) &= -2(m-1)\cot\theta P_{l,m-1}(\theta)\nonumber\\
               &= -(l+m-1)(l-m+2)P_{l,m-2}(\theta)
\end{align}
is not upwardly stable because of the $\cot\theta$ factor, which diverges when the polar angle $\theta$ is very small or very close to $\pi$ (although the polynomials themselves approach zero in these limits).

The distributions shown in Fig. \ref{fig:single} may vary significantly depending on the model. For example, Ferrers ellipsoids are finite and flat at the center, thus they do not contain the problematic particles described above and have much smaller error in single-precision. A Hernquist sphere is more representative of the general case in galaxies, being infinite and relatively centrally concentrated.

One could conceivably improve the accuracy at single-precision in several ways. In the test described above everything was calculated in single-precision, apart from some constant coefficients that were only calculated once, in double-precision, and then cast to single. It may be possible to identify the most sensitive parts of the force calculation and use double-precision just for those, or use pseudo-double-precision (as in \citealt{Nitadori09}) for part of or the entire force calculation routine.
Another possibility is to keep using single-precision for everything but prescribe special treatment to those orbits close to the $z$-axis.

\section{Discussion}

\subsection{The Methodology}

Expansion techniques, on their own, are best geared to simulate systems with a dominant single center, where it is important to minimize the effects of two-body relaxation, and where the system potential does not change radically (quickly) with time. An ideal class would be long-term secular evolution in a near-equilibrium galaxy.

Both methods presented in this work can be used to quickly calculate the gravitational potential and force on each particle in a many-body system, while discarding small scale structure. MEX comes from a Taylor-like expansion of the Green's function in the formal solution of the Poisson equation, while SCF is a Fourier-like expansion of the density. Both methods are important tools for collisionless dynamics and has been used extensively in astrophysics as discussed in the following Sections. They are comparable in terms of both accuracy and performance. In both methods, there are free parameters to be set:
\begin{enumerate}
\item Center of the expansion
\item Angular cutoff $\lmax$
\end{enumerate}
The center of the expansion could be the center of mass, but a better choice would be the bottom of the potential well. SCF has additional choices:
\begin{enumerate}
\setcounter{enumi}{2}
\item Length unit
\item Radial basis set $\{\Phi_{nl}(r)\}$
\item Radial cutoff $\nmax$
\end{enumerate}
The choice of length unit (or model scaling) affects the accuracy of SCF expansion because the zeroth order of the radial basis functions corresponds to a model of a particular scale. For example, the basis set offered by \textit{ETICS} corresponds to a \citet{Hernquist90} model with scale length $a=1$.

The main difference for the end-user is that SCF smooths the radial direction as well. This could be an advantage when $N$ is very small, since SCF will still provide a rather smooth potential, although it might not represent the real potential well at all due to random error. In MEX, particles are not completely unaware of each other, and every time two particles cross each other's shell, there is a discontinuity in the force, which may lead to large energy error when $N$ is small. This shell crossing occurs when two particles change places in the $r$-sorted list, and the particles need not be close to each other at all.

Both methods have some problems close to the center. In SCF, the limitation comes from both radial and angular expansions. The radial expansion cutoff induces a bias if the central density profile does not match the zeroth basis function, and a very non-spherical model would cause force bias at the center as well as the lobes. The latter is also a problem for MEX, which has two additional problems: the discrete nature and inevitably small number of particles when one gets arbitrarily close to the center, as well as the numerical error (and/or small step size required) due to having to calculate $1/r$ (the SCF basis function we use are completely regular at the center).

\subsection{MEX}\label{sec:discussion-mex}

It is clear from the literature that SCF has been by far more popular. But despite the above, we do not think that most authors intentionally avoided MEX, and that SCF was better publicized and became the standard. MEX is rarely used in its full form, but more frequently in the spherically symmetric version, sometimes called the ``spherical shells'' method; in this case just the monopole term is kept ($\lmax=0$). For example, it is used in the Poisson solvers of \citet{Henon75} Monte Carlo method. This hints that it might be easy to extends codes like MOCCA \citep{Giersz+08} to non-spherical cases using our version of MEX\footnote{Recently, \citet[in prep.]{Vasiliev14} introduced a new Monte Carlo code that uses SCF as a potential solver.}. This monopole approximation has also been used to study dark and stellar halo growth \citep{Lin+83, Nusser+99, Helmi+99}.

The extension of the spherical case using spherical harmonics exists in several variations, divided roughly to two classes: grid and gridless codes. The MEX version presented in this work is gridless and follows from \citet{Villumsen82} and \citet{White83}. These authors used Cartesian instead of spherical coordinates, and softened the potential at the center. This softening, albeit similar mathematically, is not equivalent to particle-particle softening in direct $N$-body simulations and was just used to prevent divergence at the center.

The first MEX code however is by \citet{Aarseth67}, who divided the simulation volume into thick shells, and the force on a particle was calculated by summing the multipoles of all shells except its own (own-shell correction was added). Similarly, \citet{Fry+80} used a MEX code with $\lmax = 3$ to explore galaxy correlation functions; in their version each shell had six particles, and softened Newtonian interaction was used within a shell. As noted in the introduction, \citet{vanAlbada+77} used a variation with axial symmetry (up to $\lmax = 4$ but with no azimuthal terms, namely $m_\mathrm{max}=0$), with a grid in both $r$ and $\theta$. in a follow up work \citep{vanAlbada82, Bontekoe87, Bertin+89, Merritt+90} the method was extended to 3D geometry. Finally, \citet{McGlynn84} used a grid in $r$ only, with logarithmic spacing. He argues that softening sacrifices the higher resolution near the center (which is one of the primary advantages of the method) and that a radial grid smooths the potential and prevents shell crossing. Recently, \citet{Vasiliev13} presented a similar potential solver with a spline instead of a grid.

We note that a virtually identical mathematical treatment to the MEX method has been applied to solve the Fokker-Planck equation under the local approximation (neglecting diffusion in position). The collisional terms of the Fokker-Planck equation can be written by means of the Rosenbluth potentials \citep{Rosenbluth+57}, which are integrals in velocity space very similar in form to equation (\ref{eq:pot-3D}). \citet{Spurzem+95} assumed azimuthal symmetry and wrote the Rosenbluth potentials using the Legendre polynomials up to $\lmax=4$ in a way exactly analogous to our equation (\ref{eq:mex-phi}). This treatment was expanded to $\lmax=5$ by \citet{Schneider+11}.

\subsection{SCF}

As noted in the previous section, SCF gained much more popularity than MEX. The SCF formalism has had wide use on galaxy-scale problems. It has been used to model the effect of black hole growth or adiabatic contraction on the structure (density profile) of the dark matter halo~\citep[e.g.][]{Sigurdsson95}. SCF is also an appropriate tool to model the growth of the stellar and dark matter halos~\citep[e.g.][]{Johnston96,Lowing+11} as well as the mass evolution of infalling satellite galaxies~\citep[e.g.][]{KHB99, KHB2000}. One of the clearest uses of the SCF technique is when the stability of the orbit matters such as in the study of chaos in galactic potentials~\citep[e.g][]{KHB01, KHB02}, and in the exchange of energy and angular momentum by mean resonances~\citep[e.g][]{WK02, KHB05}. \citet{Earn+95}  compare a number of methods and show that SCF is superior for stability work.

The initial motivation for \textit{this} work was to follow up on \citet{Meiron+12,Meiron+13}, who studied supermassive black hole binaries using a restricted technique. In their method, the stellar potential was held constant while the black holes were treated separately as collisional particles; it was thus not self-consistent in terms of the potential. This class of problems, where there
is a small subset of particles that need to be treated collisionally, has already been
attempted using an extension of the expansion technique which hybridizes SCF and direct Aarseth-type gravitational force calculation; in these extensions, either the black holes are the only collisional particle~\citep[e.g][]{Quinlan+97, Chatterjee+03}, or all centrophilic particles are treated collisionally~\citep{Hemsendorf+02}.  MEX has not been applied to this particular problem to our knowledge,
although it is as well suited as SCF.

\subsection{Implementation}

Our SCF implementation on GPU outperformed the serial Hernquist CPU version by a factor of $\sim 35$ (for double-precision), but this number depends on the particular GPU and CPU hardware compared. The CPU code is definitely competitive on multi-core CPUs. Intel recently introduced the Many Integrated Core architecture (known as Intel MIC), which are shared memory boards with the equivalent of tens of CPUs. In principle, the Fortran SCF code for CPUs could be adapted for this architecture with little modification, and it will most likely outperform the GPU version. On the other hand, next generation GPUs (such as Nvidia's Maxwell architecture) would also deliver performance improving features, and it is not clear which one would win.
The goal of this project is to ultimately enable simulations of $N \geq 10^8$, and to perform them fast enough so that many could be performed, exploring large parameter space rather than making a few such large-$N$ simulations. To do that, the code will be adapted to a multi-GPU and multi-node machines using MPI. As noted in Section \ref{sec:implementation-parallel}, this is easy for SCF but not so much for MEX.

Simultaneously we will attempt to improve the per-GPU performance. We spent a lot of time trying to optimize this first version of \textit{ETICS}, by no means we guarantee that out implementation is flawless. Some improvement might come from tweaking of the implementation. For example, we decided not to cache $P_{l0}(\cos\theta)$ but rather recalculate it in-kernel before every $m$-loop (as a starting point for the recursion relation). Since the Legendre polynomials are ``hard-coded'' and computed very efficiently, it is not immediately clear if caching is a more efficient approach (it is probably worth while at very high $\lmax$). Likewise, we chose to separate the caching operations that are performed once per routine or once per external loop, and execute them as independent kernels, while in principle they could be executed as statements inside the inner kernels (so called ``kernel fusion'', which would save kernel execution overhead), with an if-statement making sure that the cache operations are performed only if needed.

Some possible more fundamental changes include trying to get rid of the sorting operation in MEX; while the most basic approach requires the particle list to be sorted and a cumulative sum performed over the multiples, some alternatives exist such as logarithmic grid (as in \citealt{McGlynn84}) or spline (as in \citealt{Vasiliev13}). Also we might find a more sophisticated way to perform the cumulative sum, since we suspect that the \textit{Thrust} routines are not optimal for our uses. Another improvement might come from the integration side rather than force-calculations, such as implementation of higher order integrator instead of the leapfrog. Hernquist's SCF code already contains a 4th order Hermite scheme \citep{Makino91}, which is not hard to implement for GPUs, but MEX has a fundamental problem with this scheme due to shell crossing, which causes the force derivatives to be discontinuous.

\subsection{Final Remarks}

\textit{ETICS} is a powerful code, but as with any computer program, one should understand its limitation. The code in its current form should not be used for highly flattened system, or where two-body interactions are significant. The code is momentarily available upon request from the authors, but we plan to make it public, including a module to integrate it with the \textsc{amuse} framework \citep{PZ+09,Pelupessy+13}.

\acknowledgments

We thank Peter Berczik, Adi Nusser, Marcel Zemp and Eugene Vasiliev for the interesting and helpful discussions and the referee for useful comments. YM is grateful for support from the China Postdoctoral Science Foundation through grant No. 2013M530471. The special GPU accelerated supercomputer Laohu at the Center of Information and Computing at National Astronomical Observatories, Chinese Academy of Sciences, has been used for the simulations; it was funded by Ministry of Finance of People's Republic of China, under the grant ZDY Z2008-2, and by the ``Recruitment Program of Global Experts'' (Qianren) for Rainer Spurzem (2013). RS has been partially supported by NSFC (National Natural Science Foundation of China), grant No. 11073025.

\end{document}